\documentstyle[12pt]{article}
\begin{document}
\baselineskip 24pt
\newcommand{\sheptitle}
{Understanding the  Linear See-saw Neutrino Mass Relations and Large 
Mixing Angle MSW solution}
\newcommand{\shepauthor}
{N.Nimai Singh\footnote{Permanent Address: Department of Physics,
Gauhati University, Guwahati -781014, India}}

\newcommand{\shepaddress}
{ Department of Physics and Astronomy,
University of Southampton,  Southampton, 
SO17  1BJ,   U.K.}

\newcommand{\shepabstract}
 {We  study the effective ways for generating the  linear see-saw neutrino 
mass relations  and large neutrino mixing angles in two classes of 
grand unified  $SO(10)$ models where the texture of Dirac
 neutrino mass matrix is  related  to  either the charged lepton  mass 
matrix (case A) or the up-quark  mass matrix (case B). We also briefly 
analyse 
their  stability criteria and  they   are found to be stable 
under radiative corrections at low energies.}

\begin{titlepage}
\begin{flushright}
hep-ph/0009087\\
\end{flushright}
\begin{center}
{\large{\bf \sheptitle}}
\bigskip \\  \shepauthor \\ \mbox{} \\ {\it  \shepaddress} \\ \vspace{.5in}
{\bf Abstract} \bigskip \end{center} \setcounter{page}{0}
\shepabstract
\end{titlepage}

\section{Introduction}
Recent  results from Super-Kamiokande[1,2] on solar and atmospheric neutrino
 oscillations  indicate  a strong positive hint for the  existence of 
tiny neutrino masses. It has  been inferred[1] that the solution of the 
atmospheric neutrino anomaly requires a small squared mass difference between 
$\nu_{\mu}$ and $\nu_{\tau}$ with almost maximal mixing 
($\bigtriangleup m^{2}_{23}\sim(1.5-5)\times10^{-3}eV^{2}$, 
and $\sin^{2}2\theta_{23}>0.88$). Oscillation  to sterile neutrinos 
is also ruled out. In case of the solar neutrino oscillation[2],
the  small mixing angle MSW solution, the vacuum oscillation solution and 
also the oscillation to sterile neutrino, have 
been ruled out, leaving the only possible option of the large angle MSW 
solution. 
Although the solar mixing angle is relatively large, the maximal mixing is 
not allowed[2,3] ($\bigtriangleup m^{2}_{12}\sim(2.5-15)\times10^{-5}eV^{2}$, 
and $0.25<\sin^{2}2\theta_{12}<0.65$).

The above observations lead to   
at least three possible interpretations  of  the 
observed $\bigtriangleup m^{2}_{ij}=|m_{\nu i}^{2}-m_{\nu j}^{2}|,$  
in terms  of neutrino mass 
eigenvalues,viz., hierarchical, reverse hierarchical and quasi-degenerate. 
If the neutrino masses originate from the see-saw mechanism [4], then it is 
most natural to assume the existence of a physical neutrino mass hierarchy,
 though the other possibilities are not  ruled out [5]. 
For the hierarchical case the relation $m_{\nu3}>m_{\nu2}>m_{\nu1}$ implies  
$\bigtriangleup m^{2}_{23}=|m^{2}_{\nu3}-m^{2}_{\nu2}|\approx m^{2}_{\nu3}$, 
$\bigtriangleup m^{2}_{12}=|m^{2}_{\nu2}-m^{2}_{\nu1}|\approx m^{2}_{\nu2}$, 
and this fixes the ratio of the  two neutrino masses 
$m_{\nu2}/m_{\nu3}\approx \lambda^{2}$ where  $\lambda\approx 0.22$ is
 the Wolfenstein parameter[6]. 
However it does not  fix the other two neutrino mass ratio 
$m_{\nu1}/m_{\nu2}$  
from the observations. There 
are theoretical speculations[7] that this ratio
may range from $\lambda^{2}$ to $\lambda^{4}$, and approaches
  zero for massless
$m_{\nu1}$. We can see interesting  features in common
among the masses of all the fundamental fermions including neutrinos[7]:
 $$m_{b}:m_{s}:m_{d}=1:\lambda^{2}:\lambda^{4};$$   
$$m_{t}:m_{c}:m_{u}=1:\lambda^{4}:\lambda^{8}$$ 
$$m_{\tau}:m_{\mu}:m_{e}=1:\lambda^{2}:\lambda^{6};$$  
$$m_{\nu3}:m_{\nu2}:m_{\nu1}=1:\lambda^{2}:\lambda^{p}, p>2$$ 
where $p$ is  arbitrary as  neutrino mass  $m_{\nu1}$ is not yet  
fixed from the observations. 

 In the theoretical front the main  ambiguities in the see-saw mechanism[4] 
for generating small left-handed Majorana  neutrino masses,  
lie in the choice of the texture of the Dirac neutrino mass matrix $M_{\nu}$. 
Grand unified
 $SO(10)$ models  (with or without SUSY) being employed, in principle 
predict the textures of the Dirac mass matrices
$M_{u}, M_{d}, M_{e}$ along with $M_{\nu}$, and 
their group theoretical relations. The most general grand unified 
$SO(10)$ model[8] (referred to as case A) generally  predicts 
the relation of the Dirac mass matrices  $M_{u}=M_{\nu} \propto M_{d}=M_{e}$  where 
$M_{u}$ stands for 
the mass matrix of the up-type quark sector. In another class of  left-right 
symmetric models and their extension to SUSY $SO(10)$ models[9] (referred 
to as case B) one obtains the relation 
$M_{\nu}=M_{e}\tan\beta$  where $M_{e}$ is the 
charged lepton mass matrix and $\tan\beta=V_u/V_d$. 
These two cases A and B are again subject to 
further ambiguities arising from the texture  of the right-handed Majorana 
mass matrix $M_{R}$.

In the conventional quadratic see-saw mechanism[4] the tiny 
left-handed Majorana neutrino mass eigenvalues $m_{\nu_{i}}$ vary 
as Dirac mass eigenvalues  
$m_{i}^{2}$ when the right-handed Majorana neutrino masses are assumed to be 
degenerate $M_{N_{i}}=M_{N}$. This is true for pure diagonal cases[10], and 
 the neutrino mass ratios  are simply obtained as
$ m_{\nu2}/m_{\nu3}=(m^{2}_{c}/m^{2}_{t})\sim \lambda^{8}$ , and 
 $m^{2}_{\mu}/m^{2}_{\tau}\sim \lambda^{4}$ for two types of $SO(10)$
GUT models in cases A and B respectively.
Such quadratic relations  to Dirac masses  fail to conform with 
the observations[11].
Alternatively, if the eigenvalues of $M_{N_{i}}$ follow the same hierarchy as 
$m_{i}$ in case of non-degenerate right-handed neutrino mass, then it 
could be possible that $m_{\nu_{i}}$ vary linearly as $m_{i}$ 
( hence the name linear seesaw formula)[10]. 
This can be understood 
from the fact that $m_{i}=h_{i}V_{u}$ and $M_{N_{i}}=f_{i}V_{R}$, and 
one may take $f_{i}/h_{i}=(1,1,1)$ for $i=1,2,3,$ at the symmetry breaking 
scale $M_{R}$ where  the  see-saw mechanism is operative[11].
  The neutrino mass
ratio $m_{\nu2}/m_{\nu3}$ are given as  $(m_{c}/m_{t})\sim\lambda^{4}$ 
in case A, and $(m_{\mu}/m_{\tau})\sim\lambda^{2}$ in case B 
 respectively. It is quite clear that the linear see-saw mass relation with 
charged leptons in case B agrees quite well with  the neutrino mass 
relation of the 
MSW solution 
whereas the linear see-saw relation with up-quarks in case A predicts too low 
neutrino mass ratio[11]. This indicates  the ratio $f_i/h_i$ needs 
to be modified for the case A. These shortcomings had remained 
for long time untill recently Babu and Barr[12] addressed this problem in a 
class of $SO(10)$ models  where $M_{R }$ has a hierarchy similar to 
$M_{\nu}$,  and the  texture of $M_{\nu}$ is related to the 
texture of $M_{u}$ through a multiplicative factor due to Clebsch 
coefficients in analogy with  Georgi-Jarlskog  mechanism[13] in $SU(5)$ GUT.
Such modification  could rescue [12] the linear see-saw 
neutrino mass  relations in  case A , thus keeping    the 
linear see-saw mass relations for both cases A and  B
at equal footing. However such analysis is true for nearly diagonal textures, 
and the lepton mixing parameters are inherently absent[12]. 
The  desired lepton mixing angles ($\theta_{12}, \theta_{23}$) 
for both solar and atmospheric neutrino 
oscillations can in principle be generated through a 
number of ways[14]. In case of the linear see-saw formula with up-quarks 
(in case A)  it is 
easier to get large $\theta_{23}$ and small $\theta_{12}$ as in ref.[12].
 This is due to the 
fact that the large atmospheric 
mixing and small solar mixing can be imparted from the texture of 
charged lepton mass matrix, 
but it is difficult to generate large MSW solar mixing angle from charged
 lepton sector.

 In this paper we study the generation of the effective linear 
see-saw neutrino mass relations using generalised  textures  of $M_{R}$ and
 $M_{\nu}$  in both cases A and B, and also generate large solar and 
maximal atmospheric mixings. We then discuss the stability criteria of 
the linear see-saw neutrino mass relations and mixing angles under 
quantum corrections at low energies. 
\section{Models for the effective linear see-saw mass relation}
 The  left-handed Majorana neutrino mass  matrix is  
 given by the see-saw formula[4]
\begin{equation}
m_{LL}^{\nu}=-M_{\nu}M_{RR}^{-1}M_{\nu}^{T}
\end{equation}
and the  MNS mixing matrix[15] by
\begin{equation} 
V_{MNS}=V_{eL}V_{\nu L}^{\dagger}
\end{equation}
 where $V_{\nu L}$ and $V_{e L}$ are defined through the 
diagonalisation 
$m_{LL}^{diag}=V_{\nu L}m_{LL}^{\nu}V_{\nu L}^{\dagger}$, 
 and $M_{e}^{diag}=V_{eL}M_{e}V_{eR}^{\dagger}$ respectively.

We first discuss case B where there is a class of SUSY $SO(10)$ model[9] 
which predicts the relation
\begin{equation}
M_{\nu}=M_{e}\tan\beta
\end{equation} 
In the basis where the charged lepton mass matrix is diagonal
$M_{e}=Diag(m_{e},m_{\mu},m_{\tau})$, the MNS mixing matrix  in Eq.(2) 
is entirely 
from the texture of  $M_{R}$ only through the see-saw formula (1). The light Majorana 
neutrino mass 
matrix in Eq.(1) is then given by 
\begin{equation}
m_{LL}^{\nu}=-\tan^{2}\beta M_{e}M_{RR}^{-1}M_{e}
\end{equation}
Using the texture of the right-handed neutrino mass matrix[9] 
\begin{equation}
M_{R}\sim\left(\begin{array}{ccc}
         \eta & \delta & 0 \\
         \delta     & \epsilon^{2}     &  b\epsilon      \\
         0     & b\epsilon    &   1      
      \end{array}\right)v_{R}
\end{equation}
the light Majorana neutrino mass matrix in Eq.(4) becomes[9]
\begin{equation}
m_{LL}^{\nu}\sim\left(\begin{array}{ccc}
         -\frac{\epsilon^{2}}{\eta}(1-b^{2})m_{e}^{2} & \frac{\delta}{\eta}m_{e}m_{\mu} & -\frac{\delta}{\eta}b\epsilon m_{e}m_{\tau} \\
           \frac{\delta}{\eta}m_{e}m_{\mu} &  -m^{2}_{\mu}     & b\epsilon m_{\mu}m_{\tau}        \\
         -\frac{\delta}{\eta}b\epsilon m_{e}m_{\tau} & b\epsilon m_{\mu}m_{\tau}     & -(\epsilon^{2}-\frac{\delta^{2}}{\eta})m^{2}_{\tau}        
      \end{array}\right)
\end{equation}
This can  generate small angle MSW solution[9] with the proper choice of parameters in $M_{R}$
in Eq.(5). Here we are interested to generate large mixing angle MSW solution and maximal atmospheric mixing 
along with the right order of linear see-saw neutrino mass ratio.
 We take the values of the following parameters in Eq.(6):
$$\delta/\eta=-1/\lambda; b=-(1+k),$$
 $$\epsilon=\lambda^{2}; \delta=\lambda^{6}, \eta=-\lambda^{7}$$
and obtain the neutrino mass matrix (6) of the form,
\begin{equation}
m_{LL}^{\nu}\sim\left(\begin{array}{ccc}
         2k\lambda^5         & \lambda & (1+k)\lambda \\
         \lambda     &   1     &   1+k      \\
         (1+k)\lambda     &   1+k     &   1+\lambda      
      \end{array}\right)
\end{equation}
For a specific choice of the value of  $k=0.18$, Eq.(7) gives the 
neutrino mass ratio $m_{\nu2}/m_{\nu3}=0.0341\sim\lambda^{2}$,  
and the following MNS matrix,
 \begin{equation}
V_{MNS}=V_{\nu L}^\dagger= \left(\begin{array}{ccc}
         0.9210 & -0.3623 & 0.1436 \\
         -0.3607 & -0.6530 & 0.6660 \\
         0.1475     & 0.6651    &   0.7320      
      \end{array}\right)
\end{equation}
which in turn  gives large angle  solar and maximal atmospheric mixing parameters ( $\sin^{2}2\theta_{12}=0.464$ and 
 $\sin^{2}2\theta_{23}=0.991$) respectively. The neutrino mass ratio can still be increased 
with the choice of higher value of $k=0.2$.

Next we discuss the generation of  linear see-saw mass relation in case A. 
In this  class of grand unified SO(10) models[11], all the five 
Yukawa matrices $Y_{f}$ 
where $f=u,d,e,\nu,R$ are predicted by the theory. One can have  the relation, $Y_{\nu}=Y_{u}
\propto  Y_{d}=Y_{e}$,  subject to modification due to group theoretical 
Clebsch coefficients[13]. We assume  these relations are true in SUSY 
$SO(10)$ as well. The textures of these Yukawa matrices  take the following forms[11]:
$$Y_{u}\sim\left(\begin{array}{ccc}
             0  &  \lambda^{6} &  0  \\
            \lambda^{6} & \lambda^{4} & 0 \\
              0     &       0    &    1  
            \end{array}\right),
Y_{d}\sim\left(\begin{array}{ccc}
             0  &  0  & a\lambda^{3}  \\
             a\lambda^{3} & b\lambda/3 & 0 \\
             0     &  c   &   1   
            \end{array}\right),
Y_{e}\sim\left(\begin{array}{ccc}
             0 & a\lambda^{3} & 0 \\
             0 & -b\lambda & c  \\
            a\lambda^{3} & 0 & 1 
          \end{array}\right)$$
\begin{equation}
Y_{\nu}\sim\frac{3}{8}\left(\begin{array}{ccc}
              0 & \lambda^{6} & 0 \\
              \lambda^{6}  & -8\lambda^{4} & 0 \\
                0      &       0 & 1
       \end{array}\right), 
Y_{R}\sim\left(\begin{array}{ccc}
         0 & \lambda^{6} & 0 \\
         \lambda^{6}  & 0 & 0 \\
             0 & 0  & 1
        \end{array}\right)
\end{equation}
The light see-saw Majorana mass matrix (1) is given by 
\begin{equation}  
m_{LL}^{\nu}\sim\frac{9}{64}\left(\begin{array}{ccc}
               0  &  \lambda^{6} & 0  \\
               \lambda^{6} & -16\lambda^{4} & 0 \\
                 0     &     0  &  1
          \end{array}\right)
\end{equation}
The choice of the coefficients $(a,b,c)=(0.5,0.4,0.8)$  
in Eq.(9), gives good fits
to charged lepton mass hierarchy ($m_e/m_{\mu}=0.004,m_{\mu}/m_{\tau}=0.054$),
 and $m_{LL}^{\nu}$ in Eq.(10) is nearly diagonal, and gives
 the linear see-saw  neutrino mass ratios
$m_{\nu2}/m_{\nu3}=16(m_{c}/m_{t})\sim\lambda^{2}$ and 
$m_{\nu1}/m_{\nu2}=\frac{1}{(16)^{2}}(m_{u}/m_{c})\sim\lambda^{8}$.
In this model large atmospheric mixing angle is derived from charged 
lepton sector ($\sin^{2}2\theta_{23}=0.953$). However the model predicts 
SMA MSW solution ($\sin^{2}2\theta_{12}=0.013$).
In order to generate large solar mixing angle we have to modify the 
texture of the charged lepton texture in Eq.(9). Following ref.[3] one can construct 
the charged lepton texture which can give large solar mixing angle and 
maximal atmospheric angle. The most general charged lepton mass matrix in
 terms of the mass eigenvalues can be constructed using  the relation, 
 $M_{e}=V^{\dagger}_{eL}M^{diag}_{e}V_{eL}$
where we can have the input $V_{eL}$[3],
\begin{equation}
V_{eL}=\left(\begin{array}{ccc}
              -0.93 & 0.37 & 0 \\
              -0.28  & -0.70 & 0.66 \\
               0.24  & 0.61  & 0.75
       \end{array}\right)
 \end{equation}
Together with $V_{\nu L}$ extracted from $m_{LL}^{\nu}$ in Eq.(10),  
one gets  the MNS mixing matrix
\begin{equation}
V_{MNS}=\left(\begin{array}{ccc}
              -0.929 & 0.373 & 0 \\
              -0.282  & -0.699 & 0.660 \\
               0.242  & 0.609  & 0.750
       \end{array}\right), 
 \end{equation}
which gives $\sin^{2}2\theta_{23}=0.984$ and
 $\sin^{2}2\theta_{12}=0.472$ leading to maximal atmospheric mixing and 
large  solar mixings respectively. The above two examples, each for cases 
A and B, show the consistency of the 
effective linear see-saw model with the large neutrino mixings for solar 
and atmospheric neutrino oscillations.

\section{Stability under quantum corrections} 
   We  discuss  in brief the stability criteria[16]  of the linear 
see-saw mass relations under quantum corrections[17][18]. 
 The stability conditions   imply here that the ratio of two neutrino masses 
$m_{\nu2}/m_{\nu3}$ and also the two mixing
 angles $\theta_{12}$ and $\theta_{23}$ do not change much when one moves 
from the lowest right-handed neutrino mass threshold $M_{R1}\sim 10^{13}GeV$
 scale down to top-quark mass scale.  This has been found to be  true for 
mixing angles in hierarchical case [16][19].
In the diagonal charged lepton basis with $m^{diag}_{LL}=
Diag(m_{\nu1},m_{\nu2},m_{\nu3})$ and lepton mixing matrix (MNS) in Eq.(2),
\begin{equation} 
V_{MNS}=\left(\begin{array}{ccc}
                V_{e1} & V_{e2} & V_{e3}  \\
                V_{\mu1} & V_{\mu2} & V_{\mu3} \\
                V_{\tau1} & V_{\tau2} & V_{\tau3}
               \end{array}\right)
\end{equation}
the RG equation for the neutrino mass eigenvalues $m_{\nu a}$ for  MSSM 
is worked out as [20] ($t=\ln(\mu)$)
\begin{equation}
\frac{d}{dt}m_{\nu_{a}}=\frac{1}{16\pi^{2}}\Sigma_{b=e,\mu,\tau}(K+
2h_{b}^{2}V_{ba}^{2})m_{\nu_{a}}, a=1,2,3 
\end{equation}
where 
\begin{equation}
K=[-\frac{6}{5}g_{1}^{2}-6g_{2}^{2}+6TrM_{U}^{2}],
\end{equation}
Neglecting $h_{\mu}^{2}$ and $h_{e}^{2}$ in Eq.(14), and integrating 
from the lowest right-handed 
neutrino mass scale $t_{R1}=\ln(M_{R1})$ down to top-quark mass scale 
$t_{0}=\ln(m_{t})$, we get 
the mass ratio ( for $a=2,3$)
\begin{equation}
R_{23}(t_{0})/R_{23}(t_{R1})\approx\exp[2\bigtriangleup V^{2}_{\tau32}I_{\tau}]
\end{equation}
where $R_{23}=m_{\nu2}/m_{\nu3}$ and $I_{\tau}=\frac{1}{16\pi^{2}}\int^{t_{R1}}
_{t_{0}}h^{2}_{\tau}dt$. In getting Eq.(16) we have taken  
$\bigtriangleup V^{2}_{\tau32}=(V^{2}_{\tau3}-V^{2}_{\tau2})\geq 0$ 
approximately constant in the entire range of integration. For large 
$\tan\beta$ region one can take  roughly $ I_{\tau}\sim 0.15$, 
$V_{\tau3}=0.8$ 
and $V_{\tau2}
=0.6$ as in Eq.(12), then the increase in $R_{23}$ while running from $M_{R1}$ scale 
down to at $m_{t}$ scale is nearly $10\%$, thus maintaining the neutrino
mass hierarchy even at low energies[19]. This is a 
desirable result and  helps in attaining best-fit value at low energies. 
The stability of the atmospheric mixing 
parameter $S_{at}=\sin^{2}2\theta_{23}$  is clear from the 
evolution equation[20]
\begin{equation}
16\pi^{2}\frac{d}{dt}\sin^{2}2\theta_{23}=
-2\sin^{2}2\theta_{23}\cos^{2}\theta_{23}(h_{\tau}-h_{\mu})
\frac{m_{\nu3}+m_{\nu2}}{m_{\nu3}-m_{\nu2}}.
\end{equation}
Since linear see-saw neutrino mass relation guarantees hierarchical 
relation $m_{\nu3}>m_{\nu2}$,
the parameter $S_{at}$ increases at low energies as long as 
$V_{\tau2}$ approaches $V_{\tau3}$ and this helps in 
reaching maximal value at low energies. The same analysis holds true for 
the solar mixing as well[19]. 
  
\section{Conclusion}
   To summarise, we have studied the ways to generate effective 
linear see-saw neutrino mass ratios in two clasees of $SO(10)$ models 
(cases A and B) where the texture of Dirac neutrino mass matrix is 
related to either up-quark or charged lepton mass matrix. 
In both cases we  obtained   LMA  MSW solution and maximal atmospheric mixing while preserving 
linear see-saw neutrino mass relation.
It will be useful to examine more  examples in both cases  for generating LMA MSW 
solution.
 The linear  hierarchical neutrino mass ratio and the mixing angles
in these models  are found to be stable 
 under radiative corrections. Such hierarchical linear relation is important 
in finding a common dynamical scheme for generating  and 
understanding possible 
relations among  the masses of all 
fundamental fermions in nature.


\begin{thebibliography}{17}
\bibitem{ref1}H.Sobel, talk presented at the XIX International Conference 
on Neutrino Physics and Astrophysics, Sudbury, canada, June 16-21, 2000.
\bibitem{ref2}Y.Suzuki, talk presented at the XIX International Conference 
on Neutrino Physics and Astrophysics, Sudbury, canada, June 16-21, 2000.
\bibitem{ref3} B.R.Desai, U.Sarkar, A.R.Vaucher, hep-ph/0007346, and  further references therein.
\bibitem{ref4}M.Gell-Mann, P.Ramond and R.Slansky in Sanibel Talk,
CALT-68-709, Feb 1979, and in Supergravity (North Holland, Amsterdam 1979);\\
T.Yanagida in Proc.of the Workshop on Unified Theory and
Baryon Number of the Universe, KEK, Japan, 1979;\\
R.N.Mohapatra and G.Senjanovic, Phys.Rev.Lett.44(1980)912; Phys. Rev. D23 (1981) 165.
\bibitem{ref5}S.F.King and N.Nimai Singh,hep-ph/0007243, and further references therein.
\bibitem{ref6}L.Wolfenstein, Phys. Rev. Lett 51(1983) 1945.
\bibitem{ref7}P.M.Fishbane and P.Kaus, J.Phys.G:Nucl.Part.Phys.26(2000)295.
\bibitem{ref8}K.S.Babu and R.N.Mohapatra,Phys.Rev.Lett.70(1993)2845.
\bibitem{ref9}K.S.Babu, B.Dutta and R.N.Mohapatra, Phys. Rev. D60 (1999) 9500;
            Phys. Lett. B458 (1999) 93.
\bibitem{ref10}S.A.Bludman, D.C.Kennedy and  P.G.Langacker,
    Nucl. Phys. B374 (1992) 373; M.K.Parida and M.Rani, Phys. Lett. Phys. 
lett. B377 (1996) 89; M.K.Parida and N.Nimai Singh, Phys. Rev. D59 (1998) 032002.
\bibitem{ref11}N.Nimai Singh and S.Biramani Singh, Pramana  J. Phys. 54 (2000) 235;
       
S. Biramani  Singh and N. Nimai Singh,
  J. Phys. G: Nucl. Part. Phys. 25 (1999) 1009.
\bibitem{ref12}K.S.Babu and S.M.Barr,hep-ph/0004118.
\bibitem{ref13}H.Georgi and C.Jarlskog, Phys.Lett.B86(1979)297.
\bibitem{ref14}S.M.Barr and I.Dorsner,hep-ph/0003058.
\bibitem{ref15}Z.Maki, M.Nakagawa and S.Sakata, Prog.Theor.Phys.28(1962)870.
\bibitem{ref16}N.Haba and N.Okamura, Eur. Phys. J. C14 (2000) 347; 
               J. A. Casas, J. R. Espinosa, A. Ibarra and I. Navarro, 
               Nucl. Phys. B556 (1999) 3; J. Ellis, S. Lola, Phys. Lett. b458 (1999) 310; K. R. S. Balaji, A. S. Dinghe, R. N. Mohapatra, M. K. Parida,
 Phys. Rev. Lett. 84 (2000) 5034.
\bibitem{ref17}K.S.Babu, C.N.Leung and J.Pantaleone, Phys.Lett.B319(1993)319.
\bibitem{ref18}P.H.Chankowski and Z.Pluciennik, Phys.Lett.B316(1993)312.
\bibitem{ref19}S.F.King and N.Nimai Singh, hep-ph/0006229
\bibitem{ref120}P. H. Chankowski, W. Krolikowski and S. Pokorski,
 Phys. Lett. B473 (2000) 109.    
\end{thebibliography}
\end{document}